\documentstyle[]{article}
\begin{document}
\title{Frequently Asked Questions about Decoherence}
\author{Charis Anastopoulos\\
Department of Physics, University of Maryland, \\
College Park, MD20742, USA \\and\\
Spinoza Instituut,
Leuvenlaan 4, \\
3584 HE, Utrecht 
The Netherlands. \\
(present address)} 
\maketitle
{\small 
{\bf Abstract}
We give a short, critical  review of the issue of decoherence. We establish the most general framework
 in which decoherence can be discussed, how it can be quantified and  how it can be measured. 
We focus on  environment induced decoherence and its degree of usefulness for the interpretation 
of quantum theory. We finally discuss the emergence of a classical world. 
An overall   emphasis is given  in pointing at  common fallacies and misconceptions.}
\\
\\ \\
{\bf 0. The aim of this paper} \\ \\
This paper  gives a review to the phenomenon of decoherence. Its emphasis  is rather distinct than the one commonly encountered in the literature. Usually the discussion of decoherence is accompanied by an explicit or implicit acceptance of a realist  interpretational stance (usually a variation of the Everett stance). However, decoherence as a physical phenomenon is independent of the choice of interpretation and makes sense even in an operationalist perspective, like the Kopenhagen interpretation.

This review takes then a minimalist perspective: it focuses on issues that do not require any more specific commitment than that standard quantum theory is a mathematical model that 
adequately describes experimental outcomes. This is  something that all interpretational scheme  accept either either as a starting point or a consequence. We therefore refrain from entering  detailed discussion of topics and results that are heavily interpretation-dependent.

In addition, we  try to avoid  extensive discussion on
interpretational issues, but    at certain points we have to touch upon such issues, mainly 
when we find   caution about strong claims to be necessary.
\\
\\ \\ 
{\bf 1. What is meant by decoherence?} \\ \\
In full generality,  decoherence can be defined  as the phenomenon by which quantum mechanical systems 
behave as though they are 
described by {\it classical probability theory}. In other words, a quantum mechanical system 
exhibits decoherence, when all typical  features of quantum mechanical probability are suppressed. 
Since probability refers to the statistical  properties of  systems under study, decoherence   refers 
to behavior that can be inferred from the statistical analysis in a collection 
of identically prepared  systems \footnote{ Here, we shall not  consider decoherence in theories that employ 
fundamental modifications of quantum theory, 
such as adding stochastic terms in Schr\"odinger's equation \cite{collapse}. 
Our focus is on classical behavior and its emergence  from the standard quantum mechanical formalism.}.
\\ \\
Quantum theory is a theory of complex amplitudes, a  fact that is responsible for the distinctive features of quantum probability. 
In particular \\ \\
- It implies the existence of off- diagonal elements of the density matrix. Given a density matrix $\hat{\rho}$, 
the state of a system is specified not only by the probability distribution with respect to  a basis $| i \rangle $, 
$p_i = \langle i |\hat{\rho}| i \rangle $, but also by 	the off-diagonal elements $\langle i |\hat{\rho}| j \rangle$. 
The latter have no analogue in classical probability. 
\\ \\
- Interference phases appear when we study the probabilistic aspects of  quantum systems at successive instants of time.
   This behavior is highlighted by the two-slit experiment. Consider a system prepared at a state $| \psi \rangle$ 
and experimental set-up, by which there are two possible alternatives (two slits) at time $t_1$, represented by 
the projection operators   $P_1$ and $P_2$,   And  a number of possible alternatives represented by $Q_i$ at time 
$t_2 > t_1$ (the screen). The probability that the system will pass from the slit $i$ at time $t_1$ and will 
register at $j$ at time $t_2$ is equal to 
\begin{equation}
p(i, t_1; j, t_2) = \langle \psi | \hat{P}_i \hat{Q}_j \hat{P}_i | \psi \rangle 
\end{equation}
(We have assumed that the Hamiltonian is equal to zero.) If we consider the probabilities $p(j,t_2)$ that the system 
is detected in the position $j$ at time $t_2$ then we see that these probabilities do {\it not } satisfy 
the additivity condition 
\begin{equation}
p(j,t_2) = p(1,t_1;j,t_2) + p(2,t_1;j,t_2)
\end{equation}
The failure of the additivity condition to hold  is equal to $Re \langle \psi|\hat{P}_1 
\hat{Q}_j \hat{P}_2 | \psi \rangle$.
This is essentially the interference phase  between the two "histories". 
\\ \\
- General theorems (due to Bell \cite{Bell64} and Wigner \cite{Wig76}) establish that there exists 
no probability density  that can reproduce all predictions of quantum theory, in particular ones that refer 
simultaneously to non-commuting  observables.  
But even when we restrict to  commutative observables, there exists not a probability theory that can reproduce 
predictions of multi-time probabilities.
\\ \\ 
Hence, decoherence is mathematically identified  through either the diagonalisation of a density matrix, 
or the suppression of  interference phases or the ability to adequately  model a quantum system by a 
classical stochastic process \footnote{It is important to remark on a common error that arises due to the double 
semantics of the word coherence. Originally coherence referred to behavior of waves, meaning essentially 
the absence of spatial dispersion. Due to the wave nature of Schr\"odinger's equation the word was transferred there. 
However, the wave function in quantum theory is not a real wave, rather a probabilistic object. 
Hence,  quantum coherence  is fundamentally defined as a statistical concept, rather than a wave one.
In addition it always needs to refer to a particular basis.
In field theories, however,  observables
 are of a wave nature themselves. Hence, both notions of coherence can be employed. 
There is often confusion, because of this and absence
 of wave coherence is often confused with quantum decoherence. This might lead to absurd expressions,
such as {\it propagation of decoherence} or {\em local decoherence} in quantum field theoretic or many-body 
systems.
Take, for instance, the coherent states $|z  \rangle$ of the EM field. One can choose for $z ({\bf x}) $ functions, 
that correspond to classical field configurations  {\em that exhibit spatial coherence}. (The use of the word
``coherent'' 
for the name of  coherent states refers to classical coherence of the electromagnetic field).
Let us take two of them $z_1({\bf x})$ and $z_2({\bf x})$. Each state exhibits classical  coherence but  trivial 
  quantum  coherence (with respect to the phase space basis). The state $|z_1 \rangle + |z_2 \rangle$
 is highly coherent quantum mechanically, but  incoherent classically. The 
state $|z_1 + z_2 \rangle $ exhibits trivial  quantum coherence and no  classical coherence. Finally, 
the state $\frac{1}{2} (|z_1\rangle \langle z_1| + |z_2 \rangle \langle z_2|)$ exhibits neither
 classical nor quantum coherence.}.

One should nonetheless distinguish between two very different uses of the word decoherence. One 
refers to the classicalisation of the statistics of a quantum system as a  process  that takes place {\it in time }
  and the other to the emergence of classical behavior  
inherent in a sufficiently coarse-grained description of a system. 

The former notion of decoherence is the one, that  is more often found in the literature.  It is mostly 
identified with {\it environment induced decoherence} \cite{Zur1,JZ,GKJKSZ,ZuPa00}.
Typically this refers to the following situation. A system is prepared in a state
 that is a superposition of two vectors on the Hilbert space that are macroscopically 
distinct: $ | \psi \rangle = |1 \rangle + |2 \rangle $ . When we let the system evolve, 
the presence of an environment implies that the state evolves non-unitarily. Hence, the
 initial  pure state evolves into a  mixed one.  For certain types of environment, it might be the case, 
that even if   the coupling of the system to the environment is weak, the density matrix of the system evolves into  
a state that is a {\it mixture} of the macroscopically distinct states $|1 \rangle$ and $|2 \rangle $. 
In other words, the density matrix of the system becomes rapidly (approximately) diagonal in a given basis.   
For reasons that will be explained later, 
the basis in which the density matrix is diagonalised is called the {\it pointer basis} 
and the timescale after which this diagonalisation has occured is known as the {\it decoherence time}.

The other type of decoherence refers to the situation, where a coarse-grained description 
of the system can be given  in terms of classical probability theory. This is a more general
 idea of decoherence and  as such it refers to intrinsic properties of a physical system \cite{Omn94,GeHa93}. 
What we mean is the following. In general, one cannot access with perfect accuracy the
 properties of a physical system. One therefore recourses to a {\it coarse-grained} 
description at a level that us accessible to us: mathematically, this means that the
 properties of the system are described by  projection operators $P$ to subspaces 
of the  Hilbert space that are not one-dimensional ($Tr P$ quantifies the degree 
of coarse-graining). As a result of the description in term of coarse-grained 
observables, the effect of the interference phases might be suppressed. In this
 case, the system can be described by a stochastic process \cite{GeHa93,Har93a,Ana00c}. 

The environment induced decoherence is a special case of this more general 
characterisation of decoherence. If the individual quantum system is described
 by a Hilbert space $H_S$ and the environment by a Hilbert space $H_E$, the combined
 system is described by a Hilbert space $H_S \otimes H_E$. The coarse-graining 
consists in the consideration of operators only of the type $P \otimes 1$, i.e.
 ones that project only to the system's Hilbert space.
  
We shall, hence forward , refer to the environment induced decoherence as extrinsic
 (since it caused by an external agent) and the second type as intrinsic decoherence,
 since it appears as a consequence  of the basic properties of the system.
\\ \\ \\
{\bf 2.  How is decoherence quantified?}  
\\ \\ 
A naive estimation of the degree of decoherence (with respect to a basis)  comes from   the comparison of the diagonal 
to the off-diagonal elements of the density matrix of the system. 
Hence a criterion is that $|\hat{\rho}_{ij} / \hat{\rho}_{ii}| << 1$ for all $i,j$. Whenever this is true, the 
diagonal elements $\hat{\hat{\rho}}_{ii}$ define a probability distribution $p(i)$ and the basis $| i \rangle$ is a pointer 
basis for this system. 

This, however, is very imprecise. First, it is classically reasonable that for some $i$ $\hat{\rho}_{ii} = 0$.
In this case, a simple comparison with the off-diagonal elements would not be sufficient. A more 
sharp criterion can be phrased in terms of information theory \cite{Ana99}. Given a probability distribution 
$p(i)$ we can define the corresponding Shannon information as 
\begin{equation}
I[p] = - \sum_i p(i) \log p(i)
\end{equation}
In order to discuss classicality we need to compare this with a quantum mechanical information 
quantity: the von Neumann entropy 
\begin{equation}
S[\hat{\hat{\rho}}] = - Tr \hat{\rho} \log \hat{\rho}
\end{equation}
Now it is easy to verify that if we set $p(i) = \hat{\rho}_{ii}$ the following inequality holds
\begin{equation}
I[p]-S[\hat{\rho}] \geq 0 
\end{equation}
with equality only if $\hat{\rho}$ {\it is diagonal in the basis } $ | i \rangle$.

We can, therefore, consider as a  criterion for approximate diagonalisation the condition 
$I[p] - S[\hat{\rho}] << 1$. In order to establish the decoherence due to environment one has to verify that 
$I -S$ rapidly falls close to zero.

However, such criteria, that refer to a given basis are not always practical or even physically 
meaningful. First, one does not know {\it 
a priori} the basis, upon which the diagonalisation will take place, if at all. Even if this is the case there 
is no guarantee that the decoherent behaviour of the system will be present in all 
physically realised measurements. For instance, a density matrix that is approximately diagonalised 
in position, 
 can give highly non-classical results for measurements of momentum.

 One issue that is forgotten in many analyses, is that of the {\it robustness} of the pointer basis.
By this we mean, that the behaviour of the system should be "classical" not only with respect  
to the operators, that correspond to the basis, but to a larger class of them. An elementary example is 
that of a free particle ($H = p^2/2m$). In the long time-limit the state becomes approximately diagonal in the 
momentum basis, but this is  no indication of "classical behaviour".
If one starts with a superposition of two states each of them   localised in position, but with large separation of
their centers, measurements of almost all 
 observables but position would exhibit strong interference, even though  the state 
approaches (weakly) a delta function in momentum. How this is problematic 
can also be understood in light of the following remarks.

  There are many ways in which  the term  ``basis'' is used. Our 
analysis refers to discrete bases, i.e. proper orthonormal bases in the Hilbert
 space. But is studies of decoherence it is often taken to imply continuous bases 
like momentum. In this case our criteria for decoherence need to be substantially modified. 
(For instance, the corresponding 
entropies (4) are not bounded from below.) One might construct a discrete orthonormal basis by coarse 
graining a continuous one - as, for instance, von Neumann employed Gaussians to construct 
approximate position observables with discrete spectrum \cite{vNeu}. But the procedure is far from unique 
and the degree of decoherence depends on the choice of coarse graining. One would have then 
to establish a relative insensitivity to such a choice in order to 
unambiguously identify decoherence. 
For this reason,  approximate diagonalisation in a given continuous basis is not by itself 
adequate to infer that the system is effectively decoherent and additional criteria have to 
be established. One such idea is the one of the "predictability sieve" \cite{ZHP93,PZ99}, i.e. to look for bases 
consisting of states that are minimally 
entangled with the environment in the course of evolution. But generically, such states form overcomplete 
bases on the system's Hilbert space and cannot    provide by themselves an unambiguous   basis. It is  more precise to 
 talk of an "halo" of nearby pointer bases \cite{AnZu96} rather than a precise one and 
to demand approximate diagonalisation in all bases of the halo. 

It is, therefore,  necessary to employ criteria that refer to probabilistic aspects of the 
system, that are {\em not sensitive to an arbitrary choice of basis}. 
These are provided by the phase space 
description of the system, that is inherent in the structure
 of the canonical commutation relations. Since phase space observables exhaust the physical 
content of a quantum system, they can provide a most robust criterion for decoherence.
In particular, the most
useful tool in this regard is the Wigner function. This is defined as a pseudoprobability 
 distribution on the classical phase space. It is defined in terms of the density matrix $\hat{\rho}$  as 
\begin{equation}
W(q,p) = Tr ( \hat{\rho}  \hat{\Delta}(q,p))
\end{equation} 
where $\Delta(q,p)$ is an operator defined by 
\begin{equation}
\hat{\Delta}(q,p) = \int \frac{du dv}{2 \pi} e^{-iqu - i p v} e^{i u \hat{q} + i v \hat{p}}
\end{equation}

The Wigner function is not positive, hence not a true probability distribution. 
It can take also negative values. Quantum coherence manifests itself in oscillations around zero at
the scale of $\hbar$. The only pure states that give rise to positive Wigner functions are the Gaussian
ones.
For instance a  quantum state that is a superposition of two Gaussians with 
different centers, like
\begin{equation}
\psi(x) = \frac{1}{\sqrt{2}\pi^{1/4}\sigma}[ e^{- x^2/2 \sigma^2} + e^{-(x-L)^2/2 \sigma^2}] 
\end{equation}
The corresponding Wigner function is 
\begin{equation}
W(q,p)  = \frac{1}{\sqrt{2} \sigma} e^{-\frac{p^2}{\sigma^{-2}} }
\left(e^{-\frac{x^2}{\sigma^2}} + e^{- \frac{(x-L)^2}{\sigma^2}} +
2 e^{ \frac{L^2}{4 \sigma^2}}  e^{-\frac{x^2}{2 \sigma^2} - \frac{(x-L)^2}{2 \sigma^2} } \cos L p \right)
\end{equation}

The first two terms in (9) correspond to a  mixture of two classical probability distributions centered around $x = 0$ and $x = L$. The third term, however, exhibits strong oscillations and has a prefactor  that increases exponentially with the degree of separation between the two Gaussians. When $L/\sigma >1$
the Wigner function the oscillating term causes the Wigner function to take negative values. 
It is natural then to consider the suppression of such  oscillating terms  as a sign of decoherence. 
This  is equivalent to the suppression of the off-diagonal terms 
in a phase space basis (e.g. coherent states) and corresponds to  the statement that the 
quantum system can be described by a probability distribution (a positive definite Wigner function).

 The study of the Wigner function is  a  good measure for the case of environment induced decoherence. 
For the case of intrinsic decoherence, the best prescription comes from the consistent (decoherent ) histories approach to 
quantum theory \cite{Gri84,Omn88,Omn94,GeHa90,GeHa93,Har93a}. 

A history $\alpha$ is defined as a sequence of properties of the system at successive moments of time.
Hence it is represented by a sequence of projection operators $\hat{P}_{t_1}, \ldots , \hat{P}_{t_n}$. The information
about interference and probabilities is encoded in the {\it decoherence functional}, a complex valued 
function of pairs of histories. This is given by 
\begin{equation}
d(\alpha,\beta) = Tr ( \hat{C}^{\dagger}_{\alpha} \hat{\rho}_0 \hat{C}_{\beta} )
\end{equation}
where 
\begin{equation}
\hat{C}_{\alpha} = e^{i\hat{H}t_1} \hat{P}_{t_1} e^{-i\hat{H}t_1} \ldots e^{i\hat{H}
(t_n - t_{n-1})} \hat{P}_{t_n} e^{-i\hat{H}(t_n- t_{n-1})}
\end{equation}
and $\hat{\rho}_0$ the initial state of the system. The analysis of the two-slit experiment,
 we gave earlier suggests the natural following consideration. If in an exhaustive and exclusive set 
of histories, we have the property 
\begin{equation}
d(\alpha, \beta) = 0, \hspace{3cm} \alpha \neq \beta
\end{equation}
then there exists a probability measure for this set of histories, given by 
$p(\alpha) = d(\alpha, \alpha)$. This means that these histories can be described by probability
 theory. Typically, decoherent sets contain coarse-grained histories.

Hence, the construction of the decoherence functional can provide a good criterion for decoherence. However, 
the objection of robustness, can be raised in this case as in the single time description. Consistency of 
an arbitrary  set of histories might have little to do with 
classicality of a large class of observables. For this reason it is perhaps best to consider 
the decoherence functional on phase space. This can be obtained by the Wigner transform.
Such a construction is given in reference \cite{An00a} employing the techniques of continuous-time histories 
\cite{conttime}: it 
provides a natural way to determine decoherence of the most general type. 
Alternatively one can employ  information-theoretic quantities \cite{Hal93}. 
 \\ \\ \\ 
{\bf 3. How can decoherence be measured?} \\ \\
Decoherence is a probabilistic concept. Therefore, it is only in a {\it statistical} sense that we can 
talk about its presence in a physical system. In other words, it cannot make any conclusions 
about decoherence  in the 
study of an {\it individual} quantum system, since the concept does not make any operational sense there.

The only operational way of identifying decoherence lies in the consideration of the statistical behaviour 
in a {\it collection of identically prepared systems}. This means, that we need to reconstruct from
measurements (in different individual systems)  the statistical properties of the state, in which 
the ensemble is prepared,  study its evolution in time and establish the presence of decoherence, by using 
some quantitative criterion. Let us
explain 
this in some detail. First, it is {\em not
possible} to talk about decoherence by focusing on the evolution of a {\em single} variable, say $\hat{A}$. 
The reason is the following: if we write its spectral decomposition then  
\begin{equation}
\hat{A}  = \sum_i \lambda_i |i \rangle \langle i |,
\end{equation}
the expectation value will read $\langle \hat{A} \rangle = \sum_i \lambda_i \rho_{ii}$, 
in terms of the density matrix $\rho$ of the system. Clearly, measurements of $\hat{A}$ allow us to 
probe {\em only diagonal elements of the density matrix}. 
If we are to ascertain decoherent behaviour this 
is not sufficient as we would need to make statements about the off-diagonal elements as well. 

Hence, we conclude that, in order to claim that a system classicalises we need to have access to the values 
of non-commuting observables at the same time. This is not possible for a single system, however there is no 
problem when we consider ensembles of identically prepared systems. This means, that we prepare a collection 
of systems in the state $\rho$ and then at a moment of time we can perform measurements of different non-commuting 
observables at different individual systems. This way we can reconstruct the 
 statistical properties of the system in the prepared state, through standard procedures 
of data analysis and state estimation (see for example \cite{Hel,Hol}.  
Performing such a series of measurements at different moments of time, we might notice the suppression of off-diagonal  
terms in some basis, that will be a sign of decoherence. It is the author's opinion, that there is no  failproof 
way to establish that a particular physical system decoheres or not, except for the 
state estimation based on measurements of incompatible observables at successive moments of time. In many body
systems, in particular, one often confuses the wave notion of coherence with the quantum one.

This is definitely true for decoherence induced by the environment. There is a situation, however, that one can
establish decoherence of individual systems: this is the case of (approximate) determinism. In this case,
the internal dynamics of the system are such, that the coarse grained quantum mechanical observables 
are correlated according to a {\it  deterministic equation of motion}. This is believed to arise for a 
large variety of quantum systems: such would be the case of the emergence of classical mechanical laws from 
the underlying quantum theory. In that case, the existence of almost complete predictability ensures 
the effective classicality in the coarse grained description of the quantum system.
\\ \\ \\
{\bf 4. The "phenomenology" of environment induced decoherence.}
\\ \\
By phenomenology we mean the theoretical study of certain open quantum
 systems, that are thought to provide a guide for the behaviour of quantum 
systems decohering under the action of an external environment.

The first studies  emphasised the rapidity of the decoherence 
process for macroscopic systems in an incoherent (thermal) environment. A simple model by Joos and 
Zeh \cite{JZ} established a much quoted result: that even if the environment is of so low temperature 
as the cosmic microwave background, a 
superposition of two states with  difference in their centers of the order of 
1cm, for a macroscopic body (mass $m \sim 1 kg$), would lose its coherence in a time scale of the order of $10^{-23}$
seconds. In this sense, 
environment induced decoherence of macroscopically distinct superpositions  is said to be among the fastest processes
in nature \footnote{ This is valid,  of course, if we assume  that such
superpositions can be created in the first place. This  assumption is interpretation dependent, and is natural in
realist interpretations of quantum theory.However, 
the Kopemhagen interpretation or any operational scheme for quantum theory 
needs not accept  the possibility of
existence of such states,
since the macroscopic classical world is assumed, {\em a priori}.}.

The main paradigm, though, for studies of this type are the quantum Brownian motion models \cite{UnZu40}. 
They consist of a particle with mass $M$ and moving in  a potential $V(x)$ (the system), in contact  with 
a large number of harmonic oscillators  (the bath) \cite{CaLe83,GSI88,HPZ92}.
The Hamiltonian of the system is therefore 
\begin{equation}
H = \frac{p^2}{2M} + V(x) + \sum_{\alpha} (\frac{p_{\alpha}^2}{2 m_{\alpha}} 
+ \frac{1}{2} m_{\alpha} \omega_{\alpha}^2 q_{\alpha}^2 )
+ \sum_{\alpha} c_{\alpha} q_{\alpha} x 
\end{equation}
We trace out the contribution of the environment and study the evolution of the reduced density matrix. 
The contribution of the environment is 
contained in the {\it spectral density}, a function $I(\omega)$ defined by 
\begin{equation}
I(\omega) = \sum_{\alpha} \frac{c_{\alpha}^2}{2 m_{\alpha} \omega^2}
\delta(\omega - \omega_{\alpha})
\end{equation}
Three are the important  parameters that characterise  physical spectral functions: a high energy cut-off 
$\Lambda$, an exponent characterising the low-frequency behaviour and an overall multiplicative constant $\gamma$ 
that incorporates the effects of dissipation. Typically, one writes  the spectral density as
\begin{eqnarray}
I(\omega) = \gamma \omega^s e^{- \omega^2 / \Lambda^2}
\end{eqnarray}
When the initial state of the total system is assumed factorised, the density matrix of the environment is Gaussian 
and $V(x)$ is quadratic
(or zero), one can exactly solve for the propagator of the reduced density matrix of the system and construct the 
master equation, that describes its evolution. 

A well studied case is the Fokker-Planck limit: the environment is {\it ohmic} ($s=1$) and  in a thermal state with 
temperature  $T >> \Lambda$. In this case the reduced density matrix satisfies the {\it Markov} differential equation, 
which is known as the Kramers equation
\begin{equation}
i \frac{\partial}{ \partial t} \rho 
= [\frac{p^2}{2M} + V(x), \rho] - \gamma [x,\{ p, \rho \}] - 2 i M \gamma T [x,[x, \rho]].
 \end{equation}
In this regime, the evolution of a state of the form (8) shows an exponential 
suppression of the oscillating term in the corresponding Wigner function with the  decay rate of the form  of 
$e^{- 2 M \gamma T L^2 t}$, hence a decoherence time 
$t_{dec} = (M \gamma T L^2)^{-1}$. 
There are three important timescales in this model, that are generic in many open quantum systems. 

There is the inverse cut-off time $\Lambda^{-1}$, that describes the immediate  response of
 the reservoir to the quantum system.   For times $t < \Lambda^{-1}$ the factorised initial 
condition gives often a bad approximation, since at these times the evolution is very sensitive   
on high energy correlations between system and environment \cite{GSI88,HPZ92}
 that are operationally uncontrollable. There is the classicalisation time $t_{cl} = (M \gamma T )^{-1/2}$ \
cite{AnHa94}. This is an upper limit to decoherence time ; in fact one can show that after this time thermal 
fluctuations overcome the quantum ones and that the system is adequately described by the evolution of a classical 
probability distribution. This timescale governs the rate by which quantum phases move from the system to the environment.
 Finally there exists the 
relaxation time $\gamma^{-1}$, which governs the rate of energy flow from 
the system to the environment. For realistic values of the parameters, we have that
 $ \Lambda^{-1} << t_{cl} << \gamma^{-1}$. This separation of time scales conforms
 to the most clear-cut case of environment induced decoherence: the timescales 
that governs a purely quantum process (the escape of phases to the environment) 
is much smaller than the timescales of the classical energy exchange. Hence,
 even when one can consider the system as almost closed ($\gamma^{-1} \rightarrow \infty $), 
the loss of quantum phases is still important and sufficient to classicalise the system.

In other regimes, these three timescales are not widely separate. For instance, at low 
temperature (and ohmic environment) decoherence appears within a timescale of $\Lambda^{-1}$. 
However, in this regime the use of a factorised initial condition is not 
necessarily physical and one should consider the possibility that the
 decoherence phenomena predicted are artifacts of an unphysical initial condition, or at least
 that decoherence is contingent on the initial correlations of the system to the environment \cite{HPZ92}.

The Ohmic case is the standard by which to judge quantum Brownian motion. The cases of 
subohmic ($s < 1$) and supraohmic ($s > 1$) environments are different. In the former 
the response of the system to the environment is much stronger and  decoherence is more efficient, 
while the opposite behaviour is manifested in the latter case. 

There is a sense in which environment induced decoherence is highly dependent on the infrared behaviour 
of the environment. Intuitively the reason is that if the information of the quantum phases should 
leave the individual system, be spread in the environment and not return back. In order for this to happen 
it is not only necessary that the environment is {\em large}, but its recurrence time should also be large. 
So for instance, if the environment consisted of a large number of harmonic oscillators with the same frequency 
$\omega$, there would typically be a recurrence time of the order of $\omega^{-1}$ and the quantum phases would 
reappear in the quantum system after this time. (In a sense it is similar to the Poincar\'e recurrence of 
classical mechanics.) It is, therefore, essential that the recurrence time is long: in harmonic oscillator 
baths this is guaranteed by the strong presence of infra-red modes 
$\omega \rightarrow 0$ in the spectral density \cite{Kup00}.

The behaviour, we analysed  in quantum Brownian motion is often considered as paradigmatic \cite{Omn97}. 
Indeed it is a good approximation to a large class  of environments, 
   even though  it is a very special model. In addition to the spectral density,  this behaviour  is  dependent 
on the choice of the initial condition (thermal state) and the coupling between system and environment. 
There are not any studies of initial states substantially from thermal one, but it seems  reasonable that this 
classicalisation  behaviour is typical for sufficiently "classical" 
initial conditions as thermal states (or vacuum for $T = 0$) and would not persist 
in states with quantum behaviour like squeezed states.

Also, in quantum Brownian motion the system couples to the environment in the position basis.
 A resonant type of coupling (as appears for instance in atom-field interaction
\cite{AnHu00}) leads to
   decoherence in the energy basis and within a timescale, that is of the same order of
 magnitude as the relaxation time. According to our previous discussion, this is reasonable since 
a resonant type of coupling effectively selects a part of the environment's 
modes are relevant (the ones around the resonance's frequency) and misses the 
important contribution of the infra-red sector. 
It is a matter of convention whether
 one will call this type of behaviour as lying within the domain of environment
 induced decoherence, because decoherence is in a sense trivial. The flow of 
quantum phases to the environment is not distinct from the  energy flow. As such, 
the two phenomena cannot be considered as  separate.  

More general (non-linear) systems exhibit more complicated behaviour \cite{APZ96,ZuPa00}: this is a consequence of many 
time- and length- scales that characterise them. One interesting possible consequence is that decoherence 
is saturated at  a distance of separation (in the position basis). Spin baths are thought to be agents of decoherence more effective than bosonic ones at low temperature ( see \cite{Stamp} and references therein). However, such calculations involve often a perturbative expansion, which  sometimes is not a good indicator of decoherence behaviour \cite{Drag}.

In light of these remarks, we can say that in order for  a environment induced decoherence
 to be manifested in a system interacting with an environment the following requirements must 
be met \\ 
i.  The environment has to be itself in a "classical" state, like a thermal state, or a vacuum.   
Here classical refers to its behaviour with respect to its Wigner function 
description. \\ 
ii. The system-environment 
coupling  should be in a  continuous (position or momentum) basis, rather than a discrete one.
\\
iii. The spectral density of the environment has to grow slowly in the infra-red regime. 
In effect, this means that the environment responds 
more slowly in the appearance of the quantum system and hence the quantum phases are lost
 before the steady rate of energy flow commences.
\\ \\ \\  
{\bf 5. What does decoherence imply for the interpretation of quantum theory?} \\ \\
Decoherence as a phenomenon is, in general, insensitive to the interpretation one decides to 
employ for quantum theory. The transition from quantum  behaviour to classical statistics 
makes sense both in an operational setting (like the Kopenhagen interpretation) or
 a realist one. 

However, it has been historically associated with realist interpretations of quantum theory 
and more particularly with the many-worlds interpretation or its offspring. These attempt 
to interpret the quantum mechanical formalism as though it refers to individual quantum systems,
 rather than statistical ensembles. These interpretations suffer from a severe problem: 
the fact that quantum systems cannot be said to possess a given property without 
making reference to the way one reasons in order to verify the truth of this assertion.
 This is a corollary of the non-commutativity of observables in quantum theory, 
or more precisely of the non-distributivity of the lattice of propositions and is known as the Kochen-Specker's 
theorem \cite{KoSp67}. 
If one wants to talk about definite properties of an individual quantum system, 
one has always to make reference to a Hilbert space basis (or more generally an Abelian 
subalgebra of observables). And the initial focus of the many-worlds interpretation was
 to examine the presence of such bases (at least) in  measurement situations.

In a realist interpretation both the measured system and the device are described by wave
 functions, which are elements of the Hilbert spaces $H_S$ and $H_M$ respectively. 
 The Hilbert space of the combined systems is then $H_S \otimes H_M$.

A basis $| R \rangle $ in the $H_M$  is assumed to 
be associated to the measurement device, each possible $R$ corresponding to 
a different value of the pointer in the device. Also, let $|i \rangle$ 
denote a basis in $H_S$ with corresponding values $\alpha_i$ of the observable 
$\hat{A}$ for the system. 
Initially, the total system lies
 in an uncorrelated state $ | \Psi \rangle$ = $| \psi_0 \rangle \otimes | R_0 \rangle = 
(\sum_i c_i^0 |i \rangle) |R_0\rangle$ , 
with $ | \psi_0 \rangle $ the initial state of the system, $c_i^0$ its coefficients in the basis $|i \rangle $ 
 and $| R_0 \rangle$ the initial
 position of the pointer. 
As a result of the interaction the state at the end of the process will be 
$ | \Psi \rangle = \sum_{Ri} d_{iR} |i \rangle | R \rangle$. Even if there exists perfect correlation 
between initial eigenstates of the system and final values of the pointer (i.e. if $ |i \rangle |R_0 \rangle 
\rightarrow |i \rangle |R_i \rangle$, where $R_i$ is uniquely determined by the value $i$) any superposition  
in the $| i \rangle $ basis will lead to an entangled state for the total combined system. 
The apparatus is then not in a definite macroscopic configuration, which then implies that we cannot ascribe a 
property like the value of the pointer to it. This is the essence of the measurement problem, also known as the 
{\em macroobjectification } problem. One solution is the {\em infamous} wave packet reduction, proposed by von Neumann, which states that after the measurement the state of the total system reduces to a mixture \cite{vNeu} 
$\sum_i |d_{i R_i}|^2 | i \rangle \langle i | \otimes | R_i \rangle \langle R_i |$, which is interpreted in terms of classical statistics. This process is postulated in an {\em ad hoc} manner and is origin is seemingly mysterious. 

This is accompanied with another severe problem. How can we make sure that a given apparatus is 
associated to a particular observable? In other words, why would the wave packet reduction take place in the 
$|R_i \rangle $ basis, which is perfectly correlated with $i$ and not in any other? 
There seems to be an arbitrariness in the choice of the basis in which a mixture collapses. 

The environment induced decoherence has arisen as a possible solution for the second problem \cite{Zur1,Zur2} 
and is also often claimed to provide a solution to the more fundamental macroobjectification problem. 
The essential argument is that any apparatus is in contact with the environment; the environment 
induces a rapid diagonalisation of the density matrix of the apparatus in a fixed basis. 
Hence, the wave packet reduction always takes place with respect to this basis for the total system. 
This (together with the interaction Hamiltonian between measured system and apparatus) 
 is then the factor that determines, what the actual correlation between the basis of the measured system and 
the pointer basis of the apparatus. This proposal is a very natural solution to the problem of 
the arbitrariness of the pointer basis. However, there are a number of points that need to be 
addressed before this answer is taken as definite.

First, since the diagonalisation due to the environment is not exact, for a 
given state of the environment there exist a number of possible pointer bases for the 
apparatus, that are close (with respect to some natural distance in the apparatus's
 Hilbert space). It has to be shown that the correlation between the measured system and the 
apparatus is largely insensitive to the precise choice one makes for the pointer basis. 

Second, an apparatus can measure the same physical quantity in very different situations,
 that can correspond to very different states of the environment. It has therefore to be
 shown that the choice of the pointer basis is robust, within reasonable variations of the
 environment's state and constituents. This has  not yet been established from the study 
of the existing simple models.

The idea of the environment induced decoherence provide a good programme towards explaining 
the correlation of pointers in apparatuses and properties of measured systems in the realist interpretations of quantum 
theory, even though it cannot yet be taken as a definite answer. However, it is often claimed that environment induced 
decoherence provides by itself  a solution of the macroobjectification problem (e.g. \cite{Zur3}). 
This statement is not true for the following reasons:

We saw earlier that the root of the macroobjectification problem is that  the final state of the 
combined microscopic system and apparatus corresponds to no definite macroscopic superposition.
The presence of the environment would typically make the final state mixed. Let us ignore for
 the moment an immediate objection:
that there is no guarantee that the general state of the environment will allow it to play the 
role of a decohering agent  and that the resulting diagonalisation will be robust. 
There is no reason for the resulting pointer basis of the combined system to be factorisable and hence to
 correspond to macroscopically distinct properties for the quantum system. This can be explicitly 
seen in  calculations in toy detector models (one explicit calculation is 
found in \cite{Zoup}). Also there exist a theorem that
establishes that 
the combined system will  not exhibit states that are macroscopically definite under a wide variety of time 
evolution laws. Such a theorem was first proved by d' Espagnat \cite{dEsp} and recently strengthened 
by Bassi and Ghirardi \cite{BaGh00}.

One concludes therefore that environment induced decoherence cannot {\em by itself} explain the 
appearance of macroscopic definite properties and a realist interpretation of quantum theory still needs an additional postulate to account for macroobjectification, as in von Neumann's measurement theory,  the Everett stance, the collapse models or in consistent histories. 
\\ \\ \\
 {\bf 6. How is the classical world explained?} \\ \\
 One is often tempted  to explain the emergence of the classical world as a result of 
environment induced decoherence, in the sense that the definite properties that can be 
attributed to objects of our experience emerge as a result of the effect of the environment. 
However, a sufficiently classical behaviour for the environment seems to be necessary if it is to 
act as a decohering agent and we can ask what has brought the environment into such a state {\em ad-infinitum}. 
One would be then forced to employ quantum theory at increasingly large scales and at the end such a question 
can only be answered at the level of considering the question at the level of the universe \footnote{Such 
is a realist answer. For the Kopenhagen interpretation the classical world is usually taken as 
something in which the quantum description is not applicable. Strictly logically, the Kopenhagen interpretation 
does not need to explain the  emergence of the classical world; but in this case it has to admit 
the failure of quantum theory to be a universal theory. Also, how a macroscopic system of 
of quantum constituents behaves classically.}.

The issue is then raised at the level of quantum cosmology. 
Since the universe is a closed system, the reasonable way to explain classicality in this framework 
is by identifying some degrees of freedom that are all pervading and sufficiently autonomous.
There are many such proposals taking as a fundamental environment: the matter field fluctuations  \cite{Hal89}, 
the gravitational field \cite{An96}, high energy modes \cite{LoMa96}, the higher order correlation 
functions \cite{CaHu} etc. The choice is often taken according to the convenience 
of the question one wants to study. 

However, there does not exist a conclusive argument why some particular degree of freedom ought to play the degree 
of a decohering environment. In the case of the gravitational field its universality and the lack of a theory 
of quantum gravity make it a plausible candidate. Gravity is often stated as a possible cause of fundamental 
modifications in quantum theory \cite{Pen} (not necessarily of the nature of environment induced decoherence as in
\cite{An96}). 
 However, if one insists on the environment satisfying 
the laws of quantum theory one still is entitled to ask, how  come that it is in a state that is 
able to cause decoherence. The only conceivable answer is the postulation of a special initial 
condition. In addition, the question is raised, what sense does it make to talk about separation of degrees 
of freedom in highly non-linear theories, such as general relativity coupled with matter. 

An environment induced decoherence seems therefore not to be able to account convincingly about 
the emergence of classical behaviour. But perhaps classicality arises as a result of intrinsic decoherence.
This is indeed a main point of the consistent histories approach - however it does not necessitate adherence to this interpretation. To see how it works let us examine 
a simple example. Consider a system that is adequately described by Schr\"odinger's equation in one 
dimension. Its phase space would be ${\bf R}^2$. In this phase space positive operators can be 
defined that correspond to phase space cells and for sufficiently large areas of the 
enclosed cell they are close to projection operators. Each of these operators $P_C$ can be said to 
correspond to the statement that the system is with high accuracy within the phase space cell $C$.
Now, as a result of the quantum dynamics it might be that for sufficiently large cells $C$ the 
operators $P_C$ evolve as $P_{C_0} \rightarrow P_{C_t}$ with high accuracy. This implies that 
properties of the system (whenever they are definite) evolve  according to approximately deterministic 
equations of motion. This has been proved to hold for a large class of potentials \cite{Omn88,Omn94}. 
Clearly, here classicality is a result of an approximate determinism of sufficiently 
coarse-grained quantities, which appear naturally from the formalism. 

This was an idealised example for the case of a system with a single degree of freedom. One is, 
however, interested in explaining the classicality of systems that consist of a large number of degrees 
of freedom. In this case there are certain type of coarse-grainings that might effect a deterministic 
description. 

First, one might choose to focus on the evolution of {\em hydrodynamic variables}, i.e. 
variables such as energy density, particle density etc. There are some reasons  that make plausible 
that such quantities exhibit classical - in fact almost deterministic- behaviour in many-body systems.  
In particular, the densities  $\rho(x)$ of conserved quantities (e.g. energy, charge), when integrated into finite
volumes vary much slower than other currents since by virtue of the conservation equation 
$\frac{\partial \rho}{\partial t} + {\bf \nabla \cdot j} = 0$, we have
\begin{equation}
\frac{\partial}{\partial t} \int_V d^3x \rho(x) = - \int_V d^3x  {\bf \nabla \cdot j} = 
\int_{\partial V} d {\bf \sigma \cdot j},  
\end{equation}
and given a sufficiently regular volume $V$ of characteristic  length scale $l$
, the density varies with the {\em area} of its boundary, i.e. with $l^2$, 
unlike other densities that vary like $l^3$. 
So, coarse-graining in position, for sufficiently large values of $l$, 
will tend to make averaged densities of conserved quantities changing more slowly 
than other averaged densities. Given the fact, that conserved quantities trivially decohere \cite{HLM95}, this is an
indirect
suggestion that these variables would be the first to examine for effective classical behaviour. 
There are some elementary models that support this assertion \cite{Hal9899}, 
however we yet lack a conclusive general argument. In particular, it is not yet clear, whether special initial 
conditions are needed in order to guarantee the decoherence of hydrodynamic variables.

Clearly, should it be shown that hydrodynamic variables habitually decohere, this would go a long way 
towards explaining the origin of a classical world in the cosmological context first, but also for general macroscopic 
systems (e.g. rivers, planets, rocks ...). In particular , we could understand how
to
reconcile the quantum field theoretic description that is deemed necessary at the very early universe  
with the hydrodynamic/thermodynamic one, which suffices for the purposes of classical cosmology. It is the 
author's opinion that this perhaps the only unambiguous way to establish the emergence of  classicality in the 
cosmological context. 
It  is also largely insensitive to the choice of interpretative scheme for quantum theory one chooses to use. 
At this stage, however, it is fair to say that 
there is not conclusive evidence about how the classical world appears and it is likely that a special initial 
condition is needed in order to guarantee it.

\end{document}